\documentclass{amsart}

\usepackage{graphicx}
\usepackage{amssymb}

\newtheorem{lemma}{Lemma}

\newtheorem{references}{References}

\newtheorem{remarks}{Remarks}

\newcommand{\et}{{\em et al.}}

\usepackage[table]{xcolor}
\definecolor{tableShade}{HTML}{F1F5FA}
\begin{document}
\title[On weakly optimal partitions in modular networks]{On weakly optimal partitions in \\ modular networks}
\author{Jorge~R.~Busch \and~Mariano~G.~Beir\'o \and~J.~Ignacio~Alvarez-Hamelin}
\thanks{J.R. Busch, M. G. Beir\'o and J.I. Alvarez-Hamelin are with Facultad de Ingenier\'ia,
Universidad de Buenos Aires, Paseo Col\'on 850, - C1063ACV - Buenos
Aires - Argentina. E-mail: \{jbusch,mbeiro\}@fi.uba.ar,
ignacio.alvarez-hamelin@cnet.fi.uba.ar\protect\\
\indent J.I. Alvarez-Hamelin is also with INTECIN, U.B.A.
and CONICET
(Argentine Council of Scientific and Technological Research). }

\begin{abstract}
Modularity was introduced as a measure of goodness for the community
structure induced by a partition of the set of vertices
in a graph. Then, it also became an objective function used to find good
partitions, with high success. Nevertheless, some works have shown a
scaling limit and certain instabilities when finding communities with
this criterion.

Modularity has been studied proposing several formalisms, as
hamiltonians
in a Potts model or laplacians in spectral partitioning.
In this paper we present a new probabilistic formalism to analyze
modularity, and from it we derive an algorithm based on
weakly optimal partitions. This algorithm obtains good quality
partitions and also scales to large graphs.

\end{abstract}

\maketitle

\section{Introduction}
Finding communities is an important issue in complex systems, it is useful
to classify and even to predict properties in biology or
groups in sociology. A very successful method to find communities was based on {\em betweenness}~\cite{freeman:asmocbob}. This divisive clustering method led to the problem of choosing a stopping criteria.
So Newman introduced the modularity
in~\cite{newman:faecsin} and~\cite{newman:aown} as a measure of goodness of such partitions. 
This notion has shown to be rich from the theoretical viewpoint, 
and in practice it provided a unifying tool to
compare partitions obtained by a diversity of methods.
On the other hand, several methods have been devised
to obtain partitions directly by modularity optimization.
This problem has been shown to be NP-hard,
and many of the algorithms developed to approach the 
optimum are diverse adaptations of some known
algorithms for these problems,
with the notable exception of Blondel~\et~\cite{blondel:fuociln}.
From a theoretical viewpoint,
and despite the complexity problem,
modularity optimization has been shown
to have some strong limitations,
driving to partitions that
do not conform to other intuitive or formal notions
of community structure. 
These limitations are related to the scaling behavior
of modularity, that causes long correlations
in community structure, and unnatural seizes of communities.

In this paper we introduce, as in~\cite{reichardt:smocd},
a generalization of the modularity function for weighted graphs,
with a resolution parameter $t$.
We give first some properties of this generalization
analogous to known properties of the usual version.
Then we introduce a notion of weak optimality of
a partition and we study some properties
of this notion, using our tools to put new light
on some of the general limitations of modularity. We address the scaling limit problem for weakly optimal partitions, and we show some of its effects for some examples on binary trees.
Finally, we describe a fast algorithm
that gives weakly optimal partitions,
explore its similarities with~\cite{blondel:fuociln},
and compare the results with those obtained
by other means. 
The result of this comparison is rather surprising:
the values of modularity that we obtained
for standard graphs are comparable,
and in several cases better,
than those obtained by other means. 
Of course this suggests that
there is a stronger relation
between weak optimality and optimality,
explaining the performance of our algorithm
and of~\cite{blondel:fuociln} (they also obtain weakly optimal partitions).
This point deserves further investigation.

This paper is organized as follows. We introduce some probabilistic
definitions in Section~\ref{def} and we analyze the consequences in
Section~\ref{consec}. The next section presents our algorithm. We provide
proofs for the lemmas in Section~\ref{proofs}. Real complex networks are
analyzed in Section~\ref{analysis}, concluding our work in
Section~\ref{conc}.

\section{Definitions}\label{def}
\subsection{Some measures}
\label{definitions.elem}
Let $V$ be a finite set,
and $m:V\times V\rightarrow {\mathbb Z}_+$
be a non-negative integer function such that $Z=\sum_{l,r} m(l,r)>0$.
We assume throughout this work that $m$ is, in addition, symmetric,
that is $m(l,r)=m(r,l)$ for $(r,l)\in V\times V$,
and that $\sum_r m(l,r)>0$ for each $l\in V$.
Then, we consider the oriented graph $G=G(V,E)$ whose vertices are
the elements of $V$, and whose edges are the pairs $(l,r)\in V\times V$
such that $m(l,r)>0$.
That is, $G$ provided with $m$ is a weighted oriented graph,
with the property that if $(l,r)\in E$ then $(r,l)\in E$.
There  can be isolated points in $G$, but if $v$ is isolated
then $m(v,v)>0$ and there is a loop in $v$. 

We define a probability measure
$m_E$ in $V\times V$ by
\[
m_E(l,r)=\frac{m(l,r)}{Z}
\] 
and additivity. We consider the marginal probabilities defined in $V$
by
\begin{eqnarray*}
m_L(l)&=&\sum_r m_E(l,r)\\
m_R(r)&=&\sum_l m_E(l,r)
\end{eqnarray*}
and the product probability $m_{LR}$ defined
in $V\times V$ by
\[
m_{LR}(l,r)=m_L(l)m_R(r)
\]
and additivity.
Finally, for $t>0$ we shall consider the signed measure
$\mu_{t}$ in $V\times V$ given by
\[
\mu_t(S)=m_E(S)-tm_{LR}(S)
\]
for $S\subset V\times V$.
By the assumed symmetry of $m$,
we have that $m_L=m_R$, and we denote 
this marginal probability measure by $m_V$,
and $m_{LR}=m_{VV}$.
Thus
\[
\mu_t(S)=m_E(S)-tm_{VV}(S)
\]
\subsection{Partitions}
We shall consider partitions $\mathcal{C}$ of $V$,
meaning a family of pairwise disjoint not empty sets $C\subset V$
such that $\cup_{C\in\mathcal{C}} C=V$.
We shall consider the usual (lattice) partial order between
partitions of $V$,
$\mathcal C \preceq \mathcal C'$ if
$\mathcal C'$ is a refinement of $\mathcal C$,
or, which is the same, for any $C\in \mathcal C$
it holds
\[
C=\cup\mathcal C'_C
\]
where $\mathcal C'_C\doteq \{C'\in \mathcal C':C'\subset C\}$.
Notice that with this partial order,
there is always a minimal partition $\mathcal{C}_0\doteq\{V\}$
and a maximal partition $\mathcal{C}_1\doteq\{\{v\}:v\in V\}$.

Given a partition $\mathcal{C}$ of $V$,
we associate to it a set of {\em diagonal} pairs $(l,r)\in V\times V$,
by
\[
D(\mathcal{C})=\cup_{C\in\mathcal{C}} C\times C
\]
and the set of {\em off diagonal} pairs
\[
\bar D(\mathcal{C})=V\times V\setminus D(\mathcal{C})= \cup_{C,C'\in\mathcal{C},C\not= C'} C\times C'
\]

Consider a partition $\mathcal C$ of $V$,
and define $c:V\rightarrow \mathcal C$
by $c(v)=C$ if $v\in C$.
Consider then the quotient graph $G/{\mathcal C}$,
whose vertices are the elements of the partition,
with weights defined by
$m'=m/{\mathcal C}:\mathcal C\times\mathcal C\rightarrow {\mathbb Z}_+$ by 
\[
m'(C,C')=\sum_{v\in C,v'\in C'} m(v,v')
\]

Then, we obtain a signed measure $\mu'_t$ in $\mathcal C\times\mathcal C$.
Of course, if $S'\subset\mathcal C\times\mathcal C$ and 
$S=\{(v,v')\in V\times V: (c(v),c(v'))\in S'\}$, then
\[
\mu'_t(S')=\mu_t(S)
\]
\begin{remarks}
Typically $m$ will be the adjacency matrix of $G$.
If we admit more general weights in our description
it is to include in our framework this quotient graphs
and the corresponding measures. This will show to be useful
in the analysis of our algorithm, where we construct 
partitions starting from the maximal partition $\mathcal C_1$
and advancing through smaller and smaller partitions
by iteratively joining two of their elements 
(see Remarks~\ref{algorithm.remarks}).
\end{remarks}
 
\subsection{Modularity}
\begin{figure}[ht]
\includegraphics[height=0.65\textwidth,angle=-90]{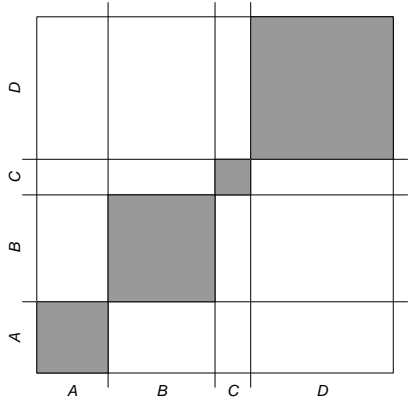}
\caption{\label{diag}Here we illustrate the set $D(\mathcal C)$ associated to a
partition $\mathcal C=\{A,B,C,D\}$. 
$V$ is plotted in the interval $[0,1]$, 
and we associate with each element $X$ in the partition an interval
of length $m_V(X)$. With this settings, you may think of $m_{VV}$
as the area, and of $m_E$ as another symmetric probability measure in
the same square.}  
\end{figure}

Now we define the modularity $Q_t(\mathcal{C})$ at resolution
$t>0$ of a partition $\mathcal C$ by
\[
Q_t(\mathcal{C})=\mu_t(D(\mathcal C))
\]
and its complement
\[
\bar Q_t(\mathcal{C})=\mu_t(\bar D(\mathcal C))
\]
(see Figure~\ref{diag})

\subsection{Optimality}
We call a partition $\mathcal C^*$ optimal for $Q_t$ when
$Q_t(\mathcal C)\leq Q_t(\mathcal C^*)$ for any other
partition $\mathcal C$.

We call a partition $\mathcal C^*$ weakly optimal for $Q_t$ when
$Q_t(\mathcal C)\leq Q_t(\mathcal C^*)$ for any 
partition $\mathcal C$ such that $\mathcal C\preceq \mathcal C^*$.

We call a partition $\mathcal C^*$ positive for $\mu_t$ when
$\mu_t(C\times C)\geq 0$ whenever 
$C$ is in $\mathcal C$.

We call a partition $\mathcal C^*$ submodular for $\mu_t$ when
$\mu_t(C\times C')\leq 0$ whenever 
$C$ and $C'$ are different sets in $\mathcal C$.
When $\mathcal C$ is submodular,
we shall call its elements communities.

We call a partition $\mathcal C$ internally connected
when $G(C)$ ({\em i.e.} the subgraph of $G$
induced by $C$) is connected for all $C\in \mathcal C$.

 
\section{Some consequences}\label{consec}
\label{conseq}
\subsection{Some useful relations}
\begin{lemma}
\label{useful}
Let $\mathcal C$ be a partition of $V$. Then
\begin{enumerate}
\item
For any $C\in \mathcal C$,
\[
\mu_t(C\times C)+\mu_t(C\times(V\setminus C))=(1-t)m_V(C) 
\]
\item
$Q_t(\mathcal C)+\bar Q_t(\mathcal C)=1-t$
\end{enumerate}
\end{lemma}

\subsection{Relations between optimality notions}
\begin{figure}[ht]
\includegraphics[height=0.65\textwidth,angle=-90]{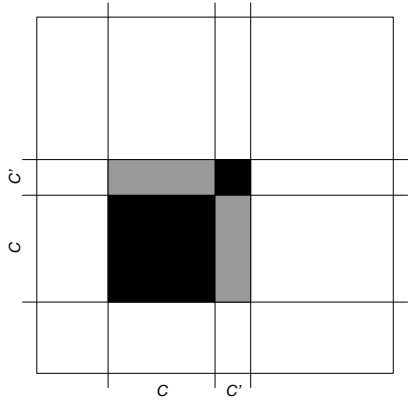}
\caption{\label{subm}
Here we illustrate Lemma~\ref{submodular.1}.
The terms associated in $Q_t(\mathcal C)$ to
$C$ and $C'$ correspond to the black squares.
When you join this sets to obtain $\mathcal D$,
you replace these two terms by one,
associated to the square formed by the black squares
and the grey rectangles. The additivity and the symmetry
of $\mu_t$ make the rest.
}  
\end{figure}

\begin{lemma}
\label{submodular.1}
Let $\mathcal C$ be a partition of $V$, and let $C,C'\in \mathcal C$
be different. Let $\mathcal D$ be the partition obtained
from $\mathcal C$ by replacing $C$ and $C'$ by $C\cup C'$,
that is
\[
\mathcal D=(\mathcal C\setminus\{C,C'\})\cup\{C\cup C'\}
\]
Then 
\[
Q_t(\mathcal D)=Q_t(\mathcal C)+2\mu_t(C\times C')
\]
\end{lemma}
(see Figure~\ref{subm})
\begin{lemma}
\label{submodular.2}
\parbox[t]{1\textwidth}{
\begin{enumerate}
\item
If $\mathcal C^*$ is optimal,
it is weakly optimal.
\item
$\mathcal C^*$ is submodular for $\mu_t$
if and only if it is weakly optimal for $Q_t$.
\item
If $\mathcal C^*$ is submodular for $\mu_t$
and $t\leq 1$, then $\mathcal C^*$ is positive for $Q_t$.
\end{enumerate}
}
\end{lemma}

\begin{lemma}
\label{conn}
Let $t>0$ and let $\mathcal C$ be any partition of $V$.
Let, for each $C\in \mathcal C$, $\mathcal D_C$ be
the partition of $C$ associated to the connected components
of $G(C)$. This defines a partition $\mathcal D$ of $V$.
Then $\mathcal D$ is internally connected
and $Q_t(\mathcal D)\geq Q_t(\mathcal C)$.
\end{lemma}

\subsection{Basic inequalities for $Q_t$}
\begin{figure}[ht]
\includegraphics[height=0.65\textwidth,angle=-90]{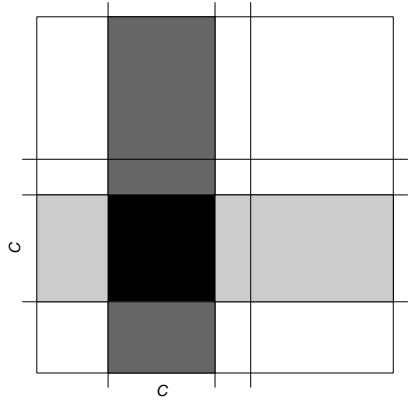}
\caption{\label{rho}
Here we illustrate Equation~\ref{rhoeq}.
The dark gray region, when you apply to it $m_E$,
gives $\rho(C)$. If you add $m_E$ applied to the
black region, you obtain $m_V(C)$ (recall that
$m_V$ is the marginal probability of $m_E$).
If you add now $m_E$ applied to the light gray region
(which is also $\rho$),
of course this, being a probability, is less than $1$. 
}  
\end{figure}

Denote 
\[
\rho(C)=m_E(C\times (V\setminus C))
\]
Then we have
\begin{lemma}
\label{bounds.set}
If $0< t$ and $C\subset V$, then
\begin{eqnarray}
m_V(C)&=&m_E(C\times C)+\rho(C)\leq m_E(C\times C)+2\rho(C) \leq 1\label{rhoeq}\\
\mu_t(C\times C)&=&m_E(C\times C)(1-t(m_E(C\times C) + 2\rho(C)))-t\rho^2(C)
\\
\mu_t(C\times C)&\leq& m_E(C\times C)(1-tm_V(C))
\\
\mu_t(C\times C)&\leq& m_E(C\times C)(1-2t\rho(C))
\end{eqnarray}
and, if in addition $t\leq 1$, 
then
\begin{eqnarray}
\mu_t(C\times C)&\geq& -t\rho^2(C)
\end{eqnarray}
\end{lemma}
(see Figure~\ref{rho})
\begin{lemma}
\label{bounds.partition}
Let $\mathcal C$ be a partition of $V$,
and $0< t$, then
\begin{eqnarray}
Q_t(\mathcal C)&\leq& 1-t\sum_{C\in \mathcal C} m_V^2(C)
\leq 1-t/|\mathcal C|\label{fixk}\\
Q_t(\mathcal C)&\leq&m_E(D(C))(1-2t\min_{C\in \mathcal C}\rho(C))
\end{eqnarray}
and, if in addition $t\leq 1$, 
then
\begin{eqnarray}
Q_t(\mathcal C)&\geq& (-t/2) (1-m_E(D(\mathcal C)))\label{lowb}
\end{eqnarray}
\end{lemma}

\begin{lemma}
\label{submodular.qbound}
Let $\mathcal C$ be a partition of $V$,
submodular for $\mu_t$. Then
\[
Q_t(\mathcal C)\geq (1-t)
\]
thus, if $t\leq 1$, $Q_t(\mathcal C)\geq 0$.
\end{lemma}

\subsection{Bounds for the size of the communities in submodular partitions: scaling limit}
\begin{lemma}
\label{bounds.k}
Let $\mathcal C$ be a partition of $V$, submodular for $\mu_t$, 
with $|\mathcal C|\geq 2$. Then
\begin{enumerate}
\item
If $C,C'\in \mathcal C$ are different, then
\begin{equation}
m_V^2(C\cup C')\geq \frac{4 m_E(C\times C')}{t}\label{scal1}
\end{equation}
\item
Assume that $G$ is connected,
let $c^*$ denote the value of the minimum cut, with weights $m$, in $G$.
Then, for all $C\in\mathcal C$ it holds
\begin{eqnarray}
\left(m_V(C)-\frac12\right)^2&\leq& \frac14 -\frac{c^*}{tZ}\\
\left(\frac{1}{|\mathcal C|}-\frac12\right)^2&\leq &\frac14 -\frac{c^*}{tZ}\\
\frac{c^*}{tZ}&<& m_V(C)< 1 -\frac{c^*}{tZ}\label{scal2}\\
|\mathcal C|&<& \frac{tZ}{c^*}\label{scal3}
\end{eqnarray}
\end{enumerate}
\end{lemma}

\subsubsection{Daisy example}
\begin{figure}[ht]
\includegraphics[height=0.65\textwidth,angle=-90]{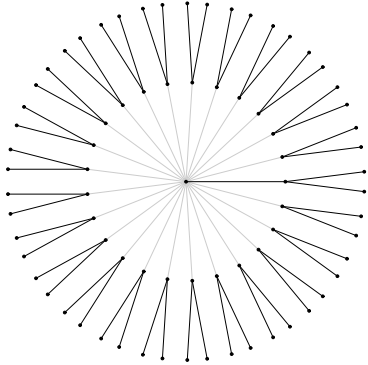}
\caption{\label{daisy2} Daisy example with $r=1$.
Here black lines represent edges
internal to a community,
and gray lines represent edges between communities.
In this case there is only one big community, 
formed up by the central vertex and one petal,
and $24$ small communities associated to the remaining petals.
}
\end{figure}

(see Figure~\ref{daisy2})
Consider a star with a center $c$ of degree $m=25r$,
and $m$ homologous $T_i$ formed by
one vertex joined to the center and two leaves.
Let $\mathcal C$ be an internally connected partition of $V$,
and assume that no element of $\mathcal C$ reduces to a leave
(see~\cite{brandes:omc}, Lemma 3.3: notice that this lemma
does {\em not} generalize to arbitrary $t>1$).
Call $C_0$ the community where $c$ lies.
Then $C_0$ is formed up by the center and $n<m$ of the $T_i$,
and the remaining elements of the partition are the remaining
$C_j=V(T_j)$.
Thus,
\[
\mu_t(C_0\times T_j)=\frac{1}{6m}\left(1-t\frac{5(m+5n)}{6m}\right)
\]

Then the pair $C_0,T_j$ is submodular when
$n\geq r(6/t-5)$. 
Let us first consider the case $t=1$. It is easy to show that you obtain
a $Q_1=\frac{4}{25}(4 - \frac{1}{6r})$ optimal partition taking $n=r$
and the remaining $24r$ $T_i$ as components.
If you increase $r$, you obtain as many
modules $T_i$ with total degree $5$ as you wish.
Of course, the number of communities
in this example, $24r+1$, 
is of the same order that $Z=150r$. 

On the other hand, we would like to add this example
to the section on counterintuitive behavior 
of modularity optimization in~\cite{brandes:omc}. 
The strong asymmetry in the community structure,
despite the strong symmetry in the graph,
and the arbitrary selection of $r$ homologous
$T_i$ for the central community,
are technical artifacts. 
This is essentially due to the presence of a
center joined to a myriad of small isolated communities,
conditions that we can not rule out from the real world.

Let $t_n=\frac{6}{5+n/r},0\leq n\leq r$ (notice that $1=t_r<\ldots<t_0=6/5$).
Then the partition $\mathcal C^*_n$ optimal for $Q_{t_n}$
has $n$ $T_i$'s in the central community,
and $m-n$ small communities $T_j$. 
This shows the influence of $t$ in the scaling limit.

\subsubsection{On complete binary trees}\label{complete_trees}
Let $G$ be a tree and let $m$ be its adjacency matrix. 
Then for any internally connected
partition $\mathcal C$ of $V$
$G/\mathcal C$ is also a tree,
and we have
\[
Q_1(\mathcal C)=1-\frac{2(|\mathcal C|-1)}{Z}-\frac{1}{|\mathcal C|}-\sum_{C\in \mathcal C} (m_V(C)-\frac{1}{|\mathcal C|})^2
\]
This follows from our definition of $Q_1$,
noticing that
\[
m_E(D(\mathcal C))=1-m_E({\bar D}(\mathcal C))=1-\frac{2(|\mathcal C|-1)}{Z}
\] 
because the number of edges between communities is, in this case,
$|\mathcal C|-1$, and that
\[
m_{VV}(D(\mathcal C))=\sum_{C\in\mathcal C} m_V^2(C)=\frac{1}{|\mathcal C|}+
\sum_{C\in\mathcal C} \left(m_V(C)-\frac{1}{|\mathcal C|}\right)^2
\]
by the well known relation between central and noncentral 
second order moments.

Let $s=|\mathcal C|$,
and consider the function 
\[
\varphi(s)=\frac{2(s-1)}{Z}+\frac{1}{s}
\]
This function has its minimum at
\[
s^*=\lfloor \frac{1+\sqrt{1+2Z}}{2} \rfloor
\]
(here $\lfloor . \rfloor$ denotes the {\em floor} function).
Of course from this we obtain the general bound for the optimal $Q_1$
of a tree
\[
Q_1^*\leq 1-\varphi(s^*)
\]

This bound is tight for complete binary trees,
because these particular graphs are almost regular,
and then the second order moment
\[
\frac{1}{s}\sum_{C\in \mathcal C} \left(m_V(C)-\frac{1}{s}\right)^2
\]
may be considered negligible. 
This is not the case for our Daisy example where we find, for $r=1$, $Q_1^*=0.613$ and $1-\varphi(s^*)=0.782$.

\begin{figure}[ht]
\includegraphics[height=0.65\textwidth,angle=-90]{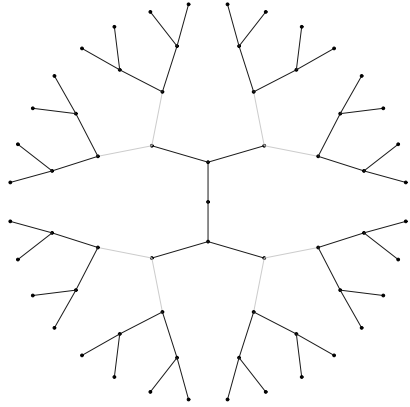}
\caption{\label{tree1}
Here we show a complete binary tree of height $5$
and its corresponding partition $\mathcal C_h$
(in this case $h=2$).
The black edges are internal to a community,
the gray ones are between communities.}  
\end{figure}

To show this,
let $G$ be a complete binary tree of height $n$,
for which $Z=2^{n+2}-4$,
let $h=\lceil (n-2)/2 \rceil$ (here $\lceil . \rceil$ stands for the {\em ceil} function)
and let us consider the partition $\mathcal C_h$ of $V$
formed by $R_h$, the vertex set of the complete binary subtree
of height $h$, and the
connected components that remain  when you remove $R_h$ from $G$ 
(see Figure~\ref{tree1}). 
Then $|\mathcal C_h|=1+2^{h+1}$.
($\mathcal C_h$ is a weakly optimal
partition, and very nearly optimal.
We shall later show some cases for which it is not optimal, see Section~\ref{trees_revisited}).
Rather than a detailed and cumbersome proof of the fact,
we show  in the following table that 
$Q_1(C_h)\approx 1-\varphi(s^*)$,
and that this approximation is better when $n$ increases.

\begin{table}[h!]
\begin{center}
\renewcommand{\arraystretch}{1.3}
\begin{tabular}{l|l|l}
$n$&$1-\varphi(s^*)$&$Q_1(\mathcal C_h)$\\
\hline
3 & 0.5357143  &0.505102\\
5&0.7620968 &0.757024 \\
6& 0.8297258 & 0.824263\\
10&0.9562724 &0.9539936 \\
20&0.9986194 & 0.998536 
\end{tabular}
\end{center}
\caption{Upper bounds and results for partitions $\mathcal C_h$}
\label{trees_upper_bounds}
\end{table}


\section{Building up submodular partitions}\label{submod}
\label{submod.build}
\subsection{Basis for an algorithm}
Let $\mathcal C$ be a partition
of $V$.
Let 
\[
t(\mathcal C)=\max 
\frac{m_E(C\times C')}{m_{VV}(C\times C')}
\] 
where $\max$ is extended to all pairs $(C,C')\in \mathcal C \times\mathcal C$
such that $C\not = C'$.
(if $|\mathcal C|=1$, we set $t(\mathcal C)=0$).
We call $t(\mathcal C)$ the resolution of $\mathcal C$. 

\begin{lemma}
\label{submod.tc}
Let $\mathcal C,\mathcal D$ be partitions of $V$ and $t>0$. Then
\begin{enumerate}
\item
$\mathcal C$ is submodular for $\mu_t$
if and only if $t\geq t(\mathcal C)$.
\item
If $\mathcal C\preceq \mathcal D$, then $t(\mathcal C)\leq t(\mathcal D)$.
\item
$t(\mathcal C)\leq t(\mathcal C_1)$
\item
$t(\mathcal C)=0$ if and only if 
$\mathcal C\preceq \mathcal B$, 
where $\mathcal B$ is the partition of $V$
associated to the connected components of $G$.
\end{enumerate}
\end{lemma}
Let $\mathcal C$ be a partition of $V$ and 
$t\geq t(\mathcal C)$. We shall use
\begin{eqnarray*}
\alpha(\mathcal C)&\doteq&m_{VV}(D(\mathcal C))=\sum_{C\in \mathcal C} m_V^2(C)\\
Z_0(\mathcal C)&\doteq &\{(C,C')\in \mathcal{C}\times\mathcal{C}, C \not = C':\mu_t(C,C')=0\}
\end{eqnarray*} 
Then we have
\begin{lemma}
\label{t.Z}
If $t\geq t(\mathcal C)>0$, then $t=t(\mathcal C)$
if and only if $Z_0(\mathcal C)\not =\emptyset$.
\end{lemma}
\begin{lemma}
\label{scheme}
Let $\mathcal C$ be a partition of $V$ 
with $t=t(\mathcal C)>0$ and let $(C,C')\in Z_0(\mathcal C)$.
Define a new partition $\mathcal D$  of $V$ by
\[
\mathcal D\doteq (\mathcal C\setminus \{C,C'\})\cup \{C\cup C'\}
\]
Then $\mathcal D\prec \mathcal C$ is submodular for $\mu_t$ and
\begin{eqnarray*}
|\mathcal D|&=&|\mathcal C|-1\\
|Z_0(\mathcal D)|&<&|Z_0(\mathcal C)|\\
Q_t(\mathcal D)&=&Q_t(\mathcal C)\\
\alpha(\mathcal D)&=&\alpha(\mathcal C)+2m_{VV}(C\times C')
\end{eqnarray*}
For $s<t$, we obtain
\[
Q_s(\mathcal D)=Q_t(\mathcal D) + (t-s)\alpha(\mathcal D) > Q_s(\mathcal C)
\] 
\end{lemma}

\begin{lemma}
\label{scheme.iter}
Let $\mathcal C$ be a partition of $V$, 
and let $t=t(\mathcal C)>0$.
Apply iteratively the scheme described in the previous lemma,
until you obtain a new partition $\mathcal D\prec C$ of $V$
such that $Z_0(\mathcal D)=\emptyset$. 

Then,
\begin{eqnarray*}
\alpha(\mathcal D)&>&\alpha(\mathcal C)\\
t(\mathcal D)&<&t\\
Q_t(\mathcal D)&=& Q_t(\mathcal C)\\ 
Q_{t(\mathcal D)}(\mathcal D)&=&
Q_{t}(\mathcal C)+
\alpha(\mathcal D)(t-t(\mathcal D))\\
\end{eqnarray*}
For $s<t$, we obtain once more
\[
Q_s(\mathcal D) > Q_s(\mathcal C)
\] 
\end{lemma}

Our algorithm is based in the last two lemmas.
Starting at $\mathcal C=\mathcal C_1$,
and $t=t(\mathcal C_1)$,
we apply iteratively the scheme described in Lemma~\ref{scheme}
until we obtain a partition $\mathcal D$ such that 
$t(\mathcal D)<t(\mathcal C)$. Now, we update $t$ to $t(\mathcal D)$,
$\mathcal C$ to $\mathcal D$, and iterate. 
The algorithm goes on while $t(\mathcal D)\geq 1$ and 
the final result is the last $\mathcal D$,
a submodular partition for $\mu_1$. 

\begin{remarks}
\label{algorithm.remarks}
After the first steps of the algorithm, we usually obtain only one partition for each resolution. Let us denote $\mathcal C_t$ to the first partition with resolution $t$. Then, $Q_t \doteq Q_t(\mathcal C_t)$ and $Q_{1t} \doteq Q_1(\mathcal C_t)$. The function $t \rightarrow Q_t$ is strictly decreasing and convex, hence $1/t \rightarrow Q_t$  is increasing and concave (see Figure~\ref{qt_graphic_2}). The function $1/t \rightarrow Q_{1t}$ is strictly increasing (see Lemma~\ref{scheme}, Lemma~\ref{scheme.iter} and Figure~\ref{qt_graphic_1}).

At the end of each step, giving a partition $\mathcal D$,
all the partitions $\mathcal D'$ considered satisfy 
$\mathcal D'\preceq \mathcal{D}$.
This means that you can update the graph to be $G/\mathcal D$ 
(doing the corresponding update in the weights), 
with a relevant gain in speed and memory.
\end{remarks}


\section{On the proofs}\label{proofs}
\subsection{Section~\ref{conseq}}
\subsubsection{Lemma~\ref{useful}}
For the first statement,
notice that both $m_E$ and $m_{VV}$ have marginal probability $m_V$.
The second statement follows from the first, adding for all $C\in\mathcal C$. 
\subsubsection{Lemma~\ref{submodular.1}}
We have already shown in Figure~\ref{subm} how this Lemma
follows, by graphical evidence, from the additivity of $\mu_t$.
\subsubsection{Lemma~\ref{submodular.2}}
\begin{enumerate}
\item
This follows immediately from the definitions.
\item
If $\mathcal C^*$ is weakly optimal,
then from Lemma~\ref{submodular.1} it follows that 
$\mu_t(C\times C')\leq 0$ for $C,C'\in \mathcal C^*,C\not =C'$,
whence $\mathcal C^*$ is submodular.

Then, if $\mathcal C^*$ is submodular and $\mathcal D\preceq \mathcal C^*$,
for any $D\in \mathcal D$
\[
\mu_t(D\times D)\leq \sum_{C\in \mathcal C^*_D}\mu_t(C\times C)
\]
and it follows that
\begin{eqnarray*}
Q_t(\mathcal D)&\leq& 
\sum_{D\in \mathcal D}\sum_{C\in \mathcal C^*_D}\mu_t(C\times C)\\
&=&Q_t(\mathcal C^*)
\end{eqnarray*}
Hence, $\mathcal C^*$ is weakly optimal.
\item
This is immediate from the first statement in our Lemma~\ref{useful}.
\end{enumerate}
\subsubsection{Lemma~\ref{conn}}
Let $C\in \mathcal C$ and $D,D'\in \mathcal D_C$.
By our definition of $\mathcal D_C$, there are no edges in $D\times D'$,
whence $m_E(D\times D')=0$ and it follows
that $\mu_t(D\times D')\leq 0$.

Then
\[
\mu_t(C\times C)\leq\sum_{D\in \mathcal D_C}\mu_t(D\times D)
\]
for all $C\in \mathcal C$, 
whence $Q_t(\mathcal C)\leq Q_t(\mathcal D)$.

\subsubsection{Lemma~\ref{bounds.set}}
In Figure~\ref{rho} we have already 
shown by graphical evidence that the first statement holds.
The second statement follows by replacing $m_V(C)$ in
$\mu_t(C\times C)=m_E(C\times C)-t m^2_V(C)$
by $m_E(C\times C)+\rho(C)$. 
The remaining statements in the Lemma are easy consequences of these two.
\subsubsection{Lemma~\ref{bounds.partition}}
\label{bounds.partition.p}
Here all is consequence of Lemma~\ref{bounds.set}.
In addition we used some general well known inequalities,
that we state here for ever:

Let $x_i,y_i$ be positive real numbers, $i=1,\ldots,n$. Then
\begin{enumerate}
\item
$\sum_i x_i^2 \geq \frac{(\sum_i x_i)^2}{n}$
\item
$\sum_i x_iy_i \leq (\max_i x_i)(\sum_i y_i)$
\end{enumerate}

\subsubsection{Lemma~\ref{submodular.qbound}}
This follows from Lemma~\ref{useful}
if you notice that, when $\mathcal C$ is submodular for $\mu_t$,
$\bar Q_t(\mathcal C)\leq 0$. 

\subsubsection{Lemma~\ref{bounds.k}}
\begin{enumerate}
\item
By the submodularity,
we have
\[
m_E(C\times C')-tm_V(C)m_V(C')\leq 0
\]
whence $m_V(C)m_V(C')\geq \frac{m_E(C\times C')}{t}$.
Now
\[
m_V^2(C\cup C')=(m_V(C)+m_V(C'))^2\geq 4m_V(C)m_V(C')
\]
and the result follows.
\item
By the submodularity, for each $C,C'\in \mathcal C$ we have
\[
m_E(C\times C')\leq tm_V(C)m_V(C')
\]
Sum for all $C'\not = C$, to obtain
\[
m_E(C\times (V\setminus C))\leq t m_V(C)(1-m_V(C))
\]
Now $Z m_E(C\times (V\setminus C))$ is a cut in $G$, whence
\[
\frac{c^*}{tZ}\leq m_V(C)(1-m_V(C))
\]
Complete squares in the right, and the first inequality follows.

As $\sum_{C\in \mathcal C} m_V(C)=1$,
we have
\[
\min_{C\in \mathcal C}m_V(C)\leq \frac{1}{|\mathcal C|}\leq\max_{C\in \mathcal C}m_V(C) 
\] 
so that the second inequality follows from the first.

From the first inequality,
we obtain
\[
|m_V(C)-\frac12|\leq\sqrt{\frac14 -\frac{c^*}{tZ}}=
\frac12\sqrt{1 -\frac{4c^*}{tZ}}
\]
whence, using the well known $\sqrt{1-x}< 1-x/2$ for $x>0$,
we obtain
\[
|m_V(C)-\frac12|\leq
\frac12 -\frac{c^*}{tZ}
\]
and the third inequality follows.
The last inequality follows from this one immediately.
\end{enumerate}

\subsection{Section~\ref{submod.build}}

\subsubsection{Lemma~\ref{submod.tc}}
\begin{enumerate}
\item
This is obvious from the definitions.
\item
If $\mathcal C\preceq \mathcal D$ and $\mathcal D$
is submodular for $\mu_t$, it follows immediately
from the additivity of $\mu_t$ that $\mathcal C$ is 
also submodular for $\mu_t$.
\item
This follows from the previous point.
\item
$t(\mathcal C)=0$ means that $m_E(C\times C')=0$
when $C,C'\in \mathcal C,C\not = C'$.
But then the connected components of $G(C)$ are, for any $C\in \mathcal C$,
connected components of $G$, whence the statement.
\end{enumerate}
\subsubsection{Lemma~\ref{t.Z}}
This follows at once from the definitions.
\subsubsection{Lemma~\ref{scheme}}
All our statements follow easily from the construction
of $\mathcal D$, perhaps with the exception of
$Z_0(\mathcal D)<Z_0(\mathcal C)$. For this,
notice that  $Z_0(\mathcal D)$
is obtained from $Z_0(\mathcal C)$
by deleting all pairs where some coordinate is $C$ or $C'$,
and adding the pairs of the form $(C\cup C',D')$ for which
$\mu_t((C\cup C')\times D')=0$, and $D'$ is neither $C$ nor $C'$. 
But $\mu_t((C\cup C')\times D')=\mu_t(C\times D')+\mu_t(C'\times D')=0$
implies that
$\mu_t(C\times D')=\mu_t (C'\times D')=0$,
so that for each pair that we eventually add,
we have deleted two 
(the same argument applies,of course, reversing the 
order in the coordinates).
As we have deleted from $Z_0(\mathcal C)$ 
at least $(C,C')$, the statement follows.

\subsubsection{Lemma~\ref{scheme.iter}}
All the statements are easy consequences
of the previous lemmas.


\section{Application to networks}\label{analysis}

We implemented our algorithm for building submodular partitions in C++; the source code is available on SourceForge~\cite{sourceforge}.

Here, we compare our results with those obtained by other methods: the
algorithm of Newman based on the spectrum of the modularity
matrix~\cite{newman:macsin}; the algorithm by Duch and Arenas using
extremal optimization~\cite{duch:cdicnueo}; the fast, greedy algorithm
of~\cite{clauset:fcsivln}; and the hierarchical fast-unfolding method
of~\cite{blondel:fuociln}. We omit previous algorithms, like the
betweenness-based Girvan-Newman method, the
spectrum-based ones and the simulated annealing method of Guimer\`{a} {\em et al.}~\cite{guimera:fcocmn},
which are rather slow and size limited. We analyze binary trees and some real networks.

\subsection{Binary trees revisited}\label{trees_revisited}

In section~\ref{complete_trees} we had found an upper bound for modularity on complete binary trees. Here, we apply our algorithm to them, and show the results in Table~\ref{table_binary_trees}, with a comparison to Blondel's algorithm~\cite{blondel:fuociln}. Both provide similar results, and quite close to the bounds in Table~\ref{trees_upper_bounds}.

\begin{figure}[ht]
\includegraphics[height=0.65\textwidth,angle=-90]{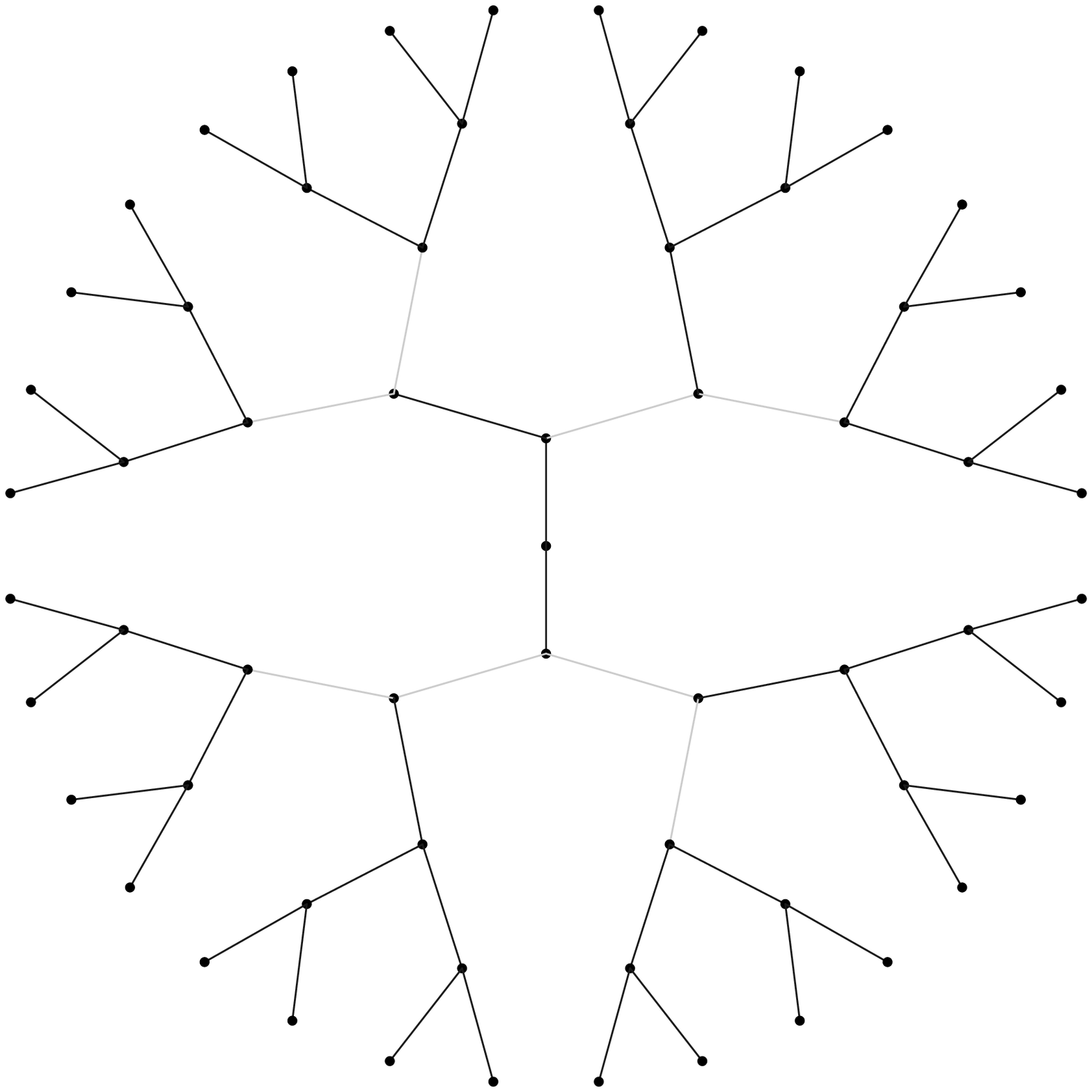}
\caption{\label{tree2}
Communities obtained by Blondel's algorithm and by this paper. We remark that for a tree of height 5, both achieve the same result.
}  
\end{figure}

We also provide a visualization of a submodular community partition for trees of height 5, in Figure~\ref{tree2}. Notice the subtle differences with the ${\mathcal C_h}$ partition in Figure~\ref{tree1}.

\begin{table}[h!]
\begin{center}
\renewcommand{\arraystretch}{1.3}
\begin{tabular}{l|l|l}
$n$ & Blondel & this paper \\
\hline
5& 0.758195 & 0.758195 \\
6& 0.821712 & 0.821051 \\
7& 0.876850 & 0.877364 \\
10& 0.953032 & 0.953219
\end{tabular}
\caption{Newman's modularity for binary trees of height $n$.}
\label{table_binary_trees}
\end{center}
\end{table}

\subsection{Real networks}\label{real_networks}

Table~\ref{resu} displays the results of $Q_1$ for different common test networks: a karate club network studied by Zachary~\cite{zachary:aifmfcafisg}, a network of email interchanges at university compiled by Guimer\`{a} {\em et al.}~\cite{guimera:sscsianohi}, a metabolic network from the {\em C. elegans}~\cite{duch:cdicnueo}, a set of scientific co-citations in arXiv~\cite{kddcup}, a trust network of users of the PGP algorithm~\cite{boguna:mosnbosda}, a coauthorship network on condensed matter physics~\cite{newman:tsoscn}, the nd.edu domain of the www~\cite{albert:tdofwww}, a web graph from Google~\cite{leskovec:csiln:ncsataolwdc} and an Internet map at the inter-router level obtained with DIMES~\cite{dimes}. This comparison table is similar to those found in~\cite{newman:macsin} and~\cite{duch:cdicnueo}.

We observe that our algorithm gives better results in terms of modularity relative to the method of Clauset {\em et al.}. For big networks, we also improve results by Newman and Duch-Arenas (not so for the smallest networks).

For larger real networks many of these methods fail, as their algorithmic complexity is too high. In those cases, we provide a comparison with the Blondel fast algorithm~\cite{blondel:fuociln}, which is also scalable and publicly available. It gives the best results for very large networks, as far as we know.

\begin{table}[ht]
\begin{center}
\renewcommand{\arraystretch}{1.3}
\rowcolors{2}{tableShade}{white}
\begin{tabular}{l|cccccc}
Network & 		Size & 	Newman &   	Duch-Arenas  &	Clauset {\em et al.} & 	Blondel & 	this paper \\
	\hline \hline
karate 		&	34   & 	0.419 &		0.419 & 	0.381 &			0.419 &		0.405 \\
dolphins   	&	62   & 	--    & 	--    & 	--    &			0.519 &		0.506 \\
email   	&	1133 & 	0.572 & 	0.574 & 	0.494 &			0.457 &		0.524 \\
metabolic	& 	453 & 	0.435 & 	0.434 & 	0.402 &			0.438 &		0.419 \\
arxiv		& 	9377 & 	--    & 	0.770 & 	0.772 &			0.813 &		0.797 \\
key signing	&	10680& 	0.855 & 	0.846 & 	0.733 &			0.884 &		0.864 \\
condmat   	& 	27519& 	0.723 & 	0.679 & 	0.668 &			0.750 &		0.723 \\
web-nd		& 	325729&	--    &		--    &		--    & 		0.935 &		0.935 \\
web-google   	&	875712&	--    &		--    &		--    &		 	0.978 &		0.968 \\
ir\_dimes   	&	976025&	--    &		--    &		--    &		 	0.845 &		0.839 \\
\end{tabular}
\end{center}
\caption{Comparison of Newman's modularity for some real networks using
different algorithms.}
\label{resu}
\end{table}

\begin{figure}[ht]
\includegraphics[height=0.75\textwidth,angle=-90]{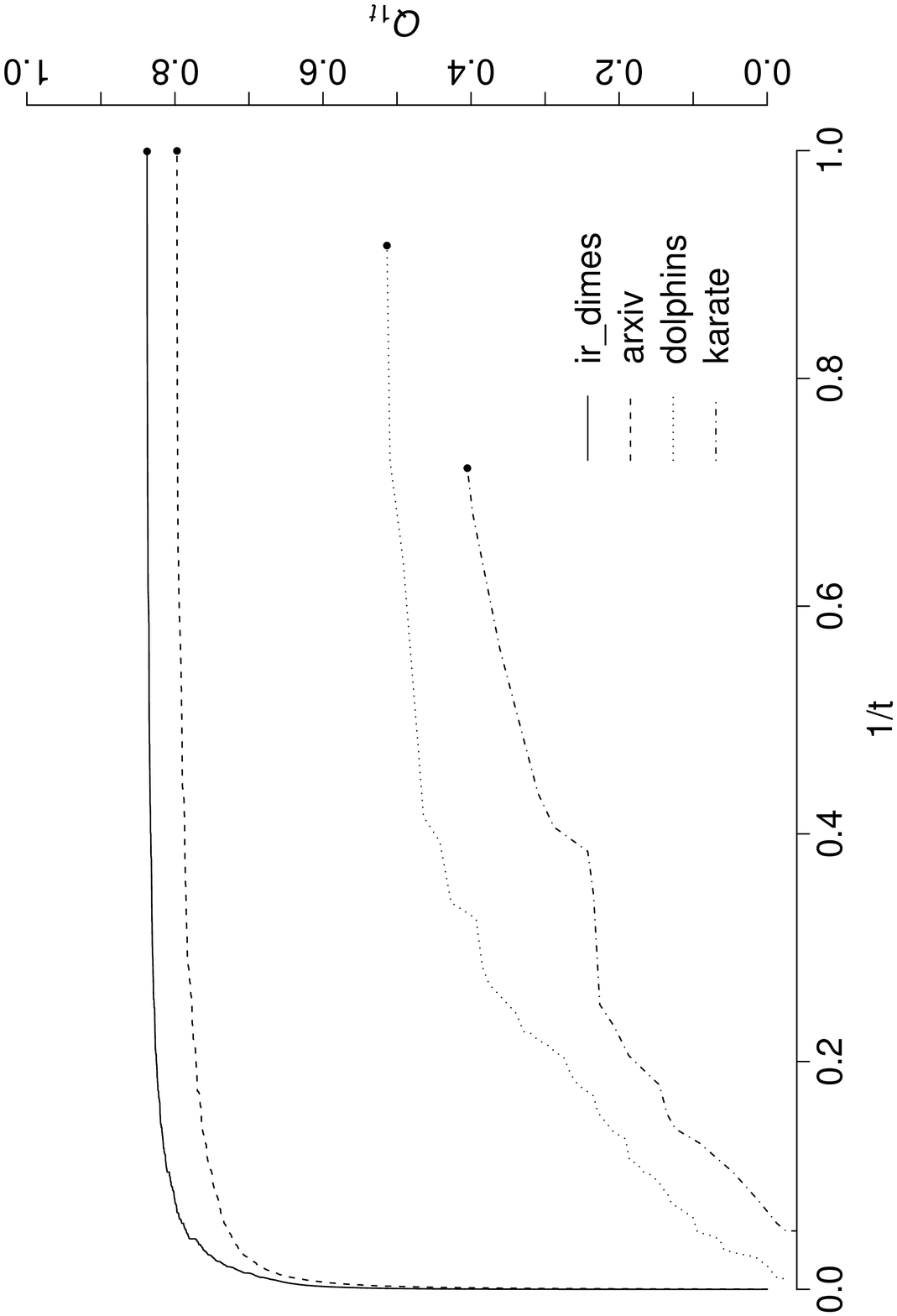}
\caption{\label{qt_graphic_1}
Evolution of $Q_1({\mathcal C_t})$ for some networks. We see that the last resolution strongly deppends on the size of the network. Even for big networks, the optimal values are reached near 1. Notice that the first few values of $Q_{1t}$ are negative; we do not plot them.}
\end{figure}
\begin{figure}[ht]
\includegraphics[height=0.75\textwidth,angle=-90]{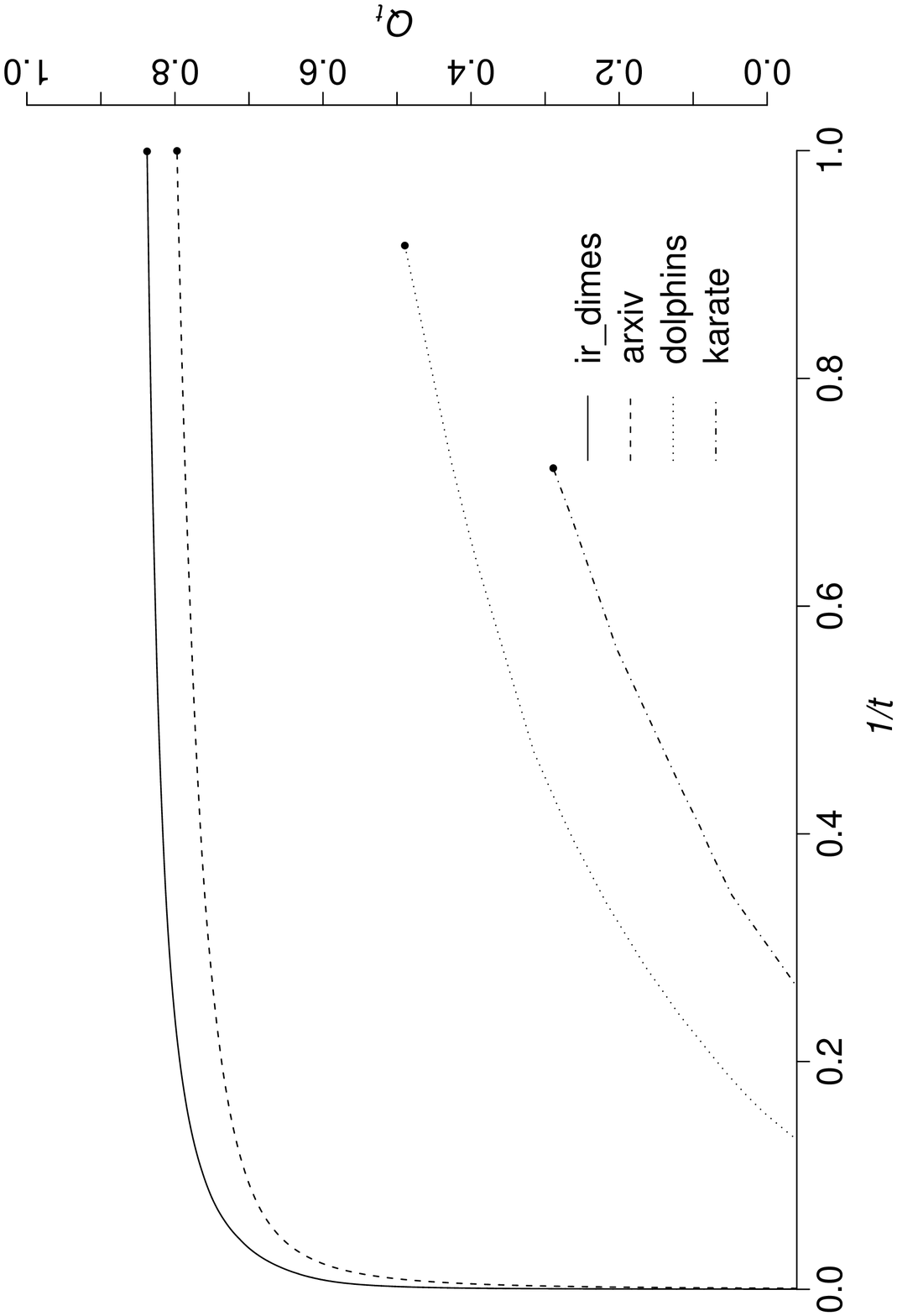}
\caption{\label{qt_graphic_2}
Evolution of $Q_t({\mathcal C_t})$ for some networks. Notice that the first values of $Q_t$ are negative. We do not plot them.}
\end{figure}

\begin{figure}[ht]
\includegraphics[height=0.75\textwidth,angle=-90]{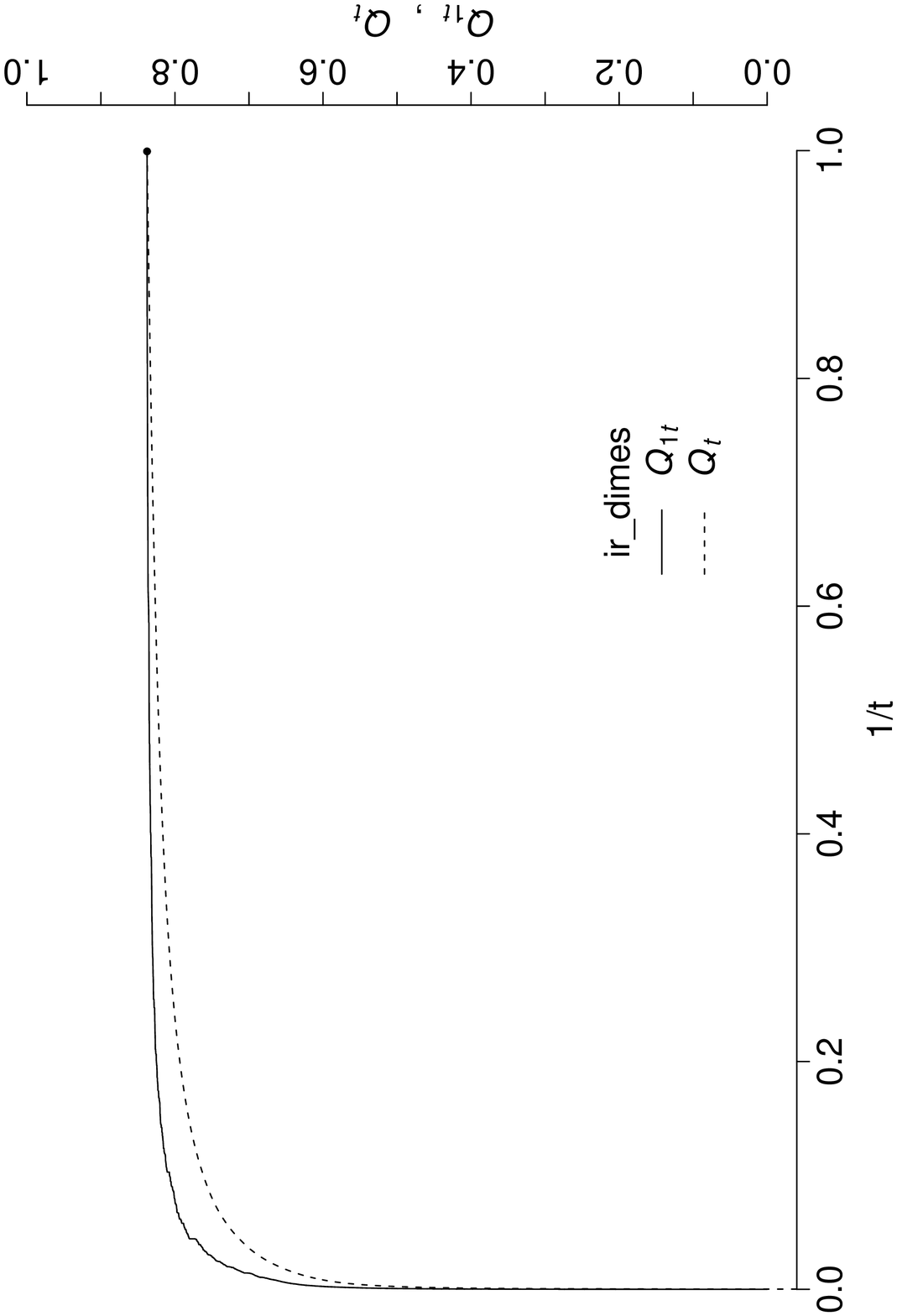}
\caption{\label{qt_graphic_3}
Comparison of $Q_t({\mathcal C_t})$ and $Q_1({\mathcal C_t})$ for the ir\_dimes network. The closeness of both curves for $t=1$ is attributable to a small second order moment of the sizes $m_v(\mathcal C)$, and to the greast number of communities.}
\end{figure}
\begin{figure}[ht]
\includegraphics[height=0.75\textwidth,angle=-90]{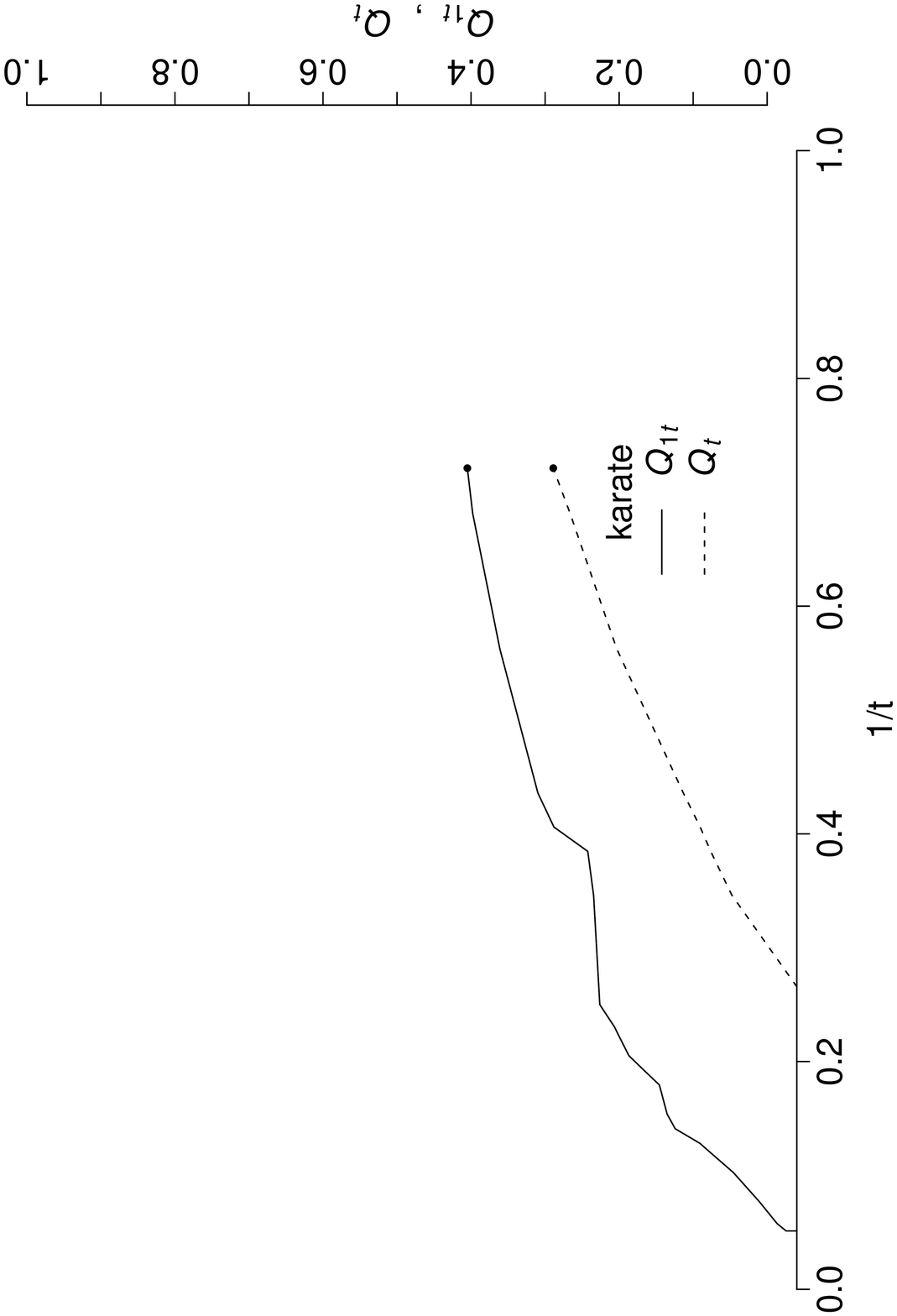}
\caption{\label{qt_graphic_4}
For karate, a small network, we see a greater contrast between $Q_1({\mathcal C_t})$ and $Q_t({\mathcal C_t})$.}
\end{figure}

To end this section, figures~\ref{qt_graphic_1} to~\ref{qt_graphic_4} display how the resolution $t$ and the modularity $Q$ evolve for different networks.

\section{Conclusions}\label{conc}

In this paper we have
shown several properties
of the modularity with resolution parameter for weighted graphs,
using systematically our version
of the definition. 
Several of these properties
were known in special cases,
as we have mentioned in detail in our reference sections;
some of them are new.
We introduced a notion of weak optimality of a partition,
and we described an algorithm to obtain weakly
optimal partitions. 
We have shown that this algorithm is able to deal with huge networks,
and that the resulting values of modularity
are comparable to those obtained by some of the
known optimization algorithms.

We showed that the known limitation
of modularity optimization,
its scaling limit, 
is also a limitation for weak optimality.
The introduction of the resolution parameter $t$
partially solves this limitation:
for $t>1$ there are weaker restrictions,
but we feel that it is necessary
to make a deeper modification in the modularity
to obtain, through its optimization,
community structures that satisfy
natural specifications.

\section*{Acknowledgments}
This work was partially funded by UBACyT I413/2008 grant. M.G.
Beir\'o acknowledges a Peruilh fellowship.

\bibliographystyle{apalike}
\bibliography{comm2.bib}

\end{document}